\documentclass[a4paper,12pt]{article}
\pdfoutput=1
\usepackage{epsfig}
\usepackage{amssymb}
\usepackage{amsfonts}
\usepackage{amsmath}
\usepackage{euscript}
\usepackage{verbatim}
\usepackage{latexsym}
\usepackage{graphicx}
\usepackage{caption}
\usepackage{float}
\usepackage{subcaption}

\usepackage{listings}

\newif\ifdtup

\catcode`\@=11

\@addtoreset{equation}{section}

\def\@normalsize{\@setsize\normalsize{15pt}\xiipt\@xiipt
\abovedisplayskip 14pt plus3pt minus3pt%
\belowdisplayskip \abovedisplayskip
\abovedisplayshortskip \z@ plus3pt%
\belowdisplayshortskip 7pt plus3.5pt minus0pt}

\def\small{\@setsize\small{13.6pt}\xipt\@xipt
\abovedisplayskip 13pt plus3pt minus3pt%
\belowdisplayskip \abovedisplayskip
\abovedisplayshortskip \z@ plus3pt%
\belowdisplayshortskip 7pt plus3.5pt minus0pt
\def\@listi{\parsep 4.5pt plus 2pt minus 1pt
     \itemsep \parsep
     \topsep 9pt plus 3pt minus 3pt}}

\relax

\catcode`@=12

\catcode`\@=11

\def\section{\@startsection{section}{1}{\z@}{3.5ex plus 1ex minus
   .2ex}{2.3ex plus .2ex}{\large\bf}}

\def\SymBoxes#1#2#3#4{\newdimen\un@t \un@t#3%
\raisebox{#1}{\rule{#2\un@t}{#4}\hskip-#2\un@t
\@tempdimb\un@t \advance\@tempdimb by-#4\@tempcntb#2\relax%
\@whilenum{\@tempcntb>0}\do{
\rule{#4}{\un@t}\hskip\@tempdimb \advance\@tempcntb by\m@ne}%
\hskip-#2\un@t \rule[\un@t]{#2\un@t}{#4}%
\rule[\un@t]{#4}{#4}\hskip-#4
\rule{#4}{\un@t}}\hskip-#4}                

\begin{document}

\newcommand{\beq}{\begin{equation}}
\newcommand{\eeq}{\end{equation}}
\newcommand{\bea}{\begin{eqnarray}}
\newcommand{\eea}{\end{eqnarray}}
\newcommand{\beas}{\begin{eqnarray*}}
\newcommand{\eeas}{\end{eqnarray*}}
\newcommand{\defi}{\stackrel{\rm def}{=}}
\newcommand{\non}{\nonumber}
\newcommand{\bquo}{\begin{quote}}
\newcommand{\enqu}{\end{quote}}
\renewcommand{\(}{\begin{equation}}
\renewcommand{\)}{\end{equation}}
\def \eqn#1#2{\begin{equation}#2\label{#1}\end{equation}}

\def\e{\epsilon}
\def\IZ{{\mathbb Z}}
\def\IR{{\mathbb R}}
\def\IC{{\mathbb C}}
\def\IQ{{\mathbb Q}}
\def\IH{{\mathbb H}}
\def\de{\partial}
\def\Tr{ \hbox{\rm Tr}}
\def\H{ \hbox{\rm H}}
\def\HE{ \hbox{$\rm H^{even}$}}
\def\HO{ \hbox{$\rm H^{odd}$}}
\def\K{ \hbox{\rm K}}
\def\Im{ \hbox{\rm Im}}
\def\Ker{ \hbox{\rm Ker}}
\def\const{\hbox {\rm const.}}
\def\o{\over}
\def\im{\hbox{\rm Im}}
\def\re{\hbox{\rm Re}}
\def\bra{\langle}\def\ket{\rangle}
\def\Arg{\hbox {\rm Arg}}
\def\Re{\hbox {\rm Re}}
\def\Im{\hbox {\rm Im}}
\def\exo{\hbox {\rm exp}}
\def\diag{\hbox{\rm diag}}
\def\longvert{{\rule[-2mm]{0.1mm}{7mm}}\,}
\def\a{\alpha}
\def\dag{{}^{\dagger}}
\def\tq{{\widetilde q}}
\def\p{{}^{\prime}}
\def\W{W}
\def\N{{\cal N}}
\def\hsp{,\hspace{.7cm}}

\def\br{\nonumber}
\def\IZ{{\mathbb Z}}
\def\IR{{\mathbb R}}
\def\IC{{\mathbb C}}
\def\IQ{{\mathbb Q}}
\def\IP{{\mathbb P}}
\def \eqn#1#2{\begin{equation}#2\label{#1}\end{equation}}

\newcommand{\C}{\ensuremath{\mathbb C}}
\newcommand{\Z}{\ensuremath{\mathbb Z}}
\newcommand{\R}{\ensuremath{\mathbb R}}
\newcommand{\rp}{\ensuremath{\mathbb {RP}}}
\newcommand{\cp}{\ensuremath{\mathbb {CP}}}
\newcommand{\vac}{\ensuremath{|0\rangle}}
\newcommand{\vact}{\ensuremath{|00\rangle}                    }
\newcommand{\oc}{\ensuremath{\overline{c}}}
\newcommand{\psizero}{\psi_{0}}
\newcommand{\phizero}{\phi_{0}}
\newcommand{\hzero}{h_{0}}
\newcommand{\psiin}{\psi_{\rh}}
\newcommand{\phiin}{\phi_{\rh}}
\newcommand{\hin}{h_{\rh}}
\newcommand{\rh}{r_{h}}
\newcommand{\rb}{r_{S}}
\newcommand{\psibnd}{\psi_{0}^{b}}
\newcommand{\psibndp}{\psi_{1}^{b}}
\newcommand{\phibnd}{\phi_{0}^{b}}
\newcommand{\phibndp}{\phi_{1}^{b}}
\newcommand{\gbnd}{g_{0}^{b}}
\newcommand{\hbnd}{h_{0}^{b}}
\newcommand{\zh}{z_{h}}
\newcommand{\zb}{z_{S}}
\newcommand{\man}{\mathcal{M}}
\newcommand{\hbr}{\bar{h}}
\newcommand{\tbr}{\bar{t}}

\begin{titlepage}
\begin{flushright}
CHEP XXXXX
\end{flushright}
\bigskip
\def\thefootnote{\fnsymbol{footnote}}

\begin{center}
{\large 
{\bf Dirichlet Baths and the Not-so-Fine-Grained Page Curve
}
}
\end{center}

\bigskip
\begin{center}
Kausik GHOSH$^a$\footnote{\texttt{kau.rock91@gmail.com}}, \ \ 
Chethan KRISHNAN$^a$\footnote{\texttt{chethan.krishnan@gmail.com}},
\vspace{0.1in}

\end{center}

\renewcommand{\thefootnote}{\arabic{footnote}}

\begin{center}

$^a$ {Center for High Energy Physics,\\
Indian Institute of Science, Bangalore 560012, India}\\

\end{center}

\noindent
\begin{center} {\bf Abstract} \end{center}

We present a doubly holographic prescription for computing entanglement entropy on a gravitating brane. It involves a Ryu-Takayanagi surface with a Dirichlet anchoring condition. In braneworld cosmology, a related approach was used previously in arXiv:2007.06551. There, the prescription naturally computed a co-moving entanglement entropy, and was argued to resolve the information paradox for a black hole living in the cosmology. In this paper, we show that the Dirichlet prescription leads to reasonable results, when applied to a recently studied wedge holography set up with a gravitating bath. The nature of the information paradox and its resolution in our Dirichlet problem have a natural understanding in terms of the strength of gravity on the two branes and at the anchoring location. By sliding the anchor to the defect, we demonstrate that the limit where gravity decouples from the anchor is continuous -- in other words, as far as island physics is considered, weak gravity on the anchor is identical to no gravity. The weak and (moderately) strong gravity regions on the brane are separated by a ``Dirichlet wall". We find an intricate interplay between various extremal surfaces, with an island coming to the rescue whenever there is an information paradox. This is despite the presence of massless gravitons in the spectrum. The overall physics is consistent with the slogan that gravity becomes ``more holographic", as it gets stronger. Our observations strengthen the case that the conventional Page curve is indeed of significance, when discussing the information paradox in flat space. We work in high enough dimensions so that the graviton is non-trivial, and our results are in line with the previous discussions on gravitating baths in arXiv:2005.02993 and arXiv:2007.06551.

\vspace{1.6 cm}
\vfill

\end{titlepage}

\setcounter{footnote}{0}

\section{Introduction}

That there are no truly local observables in the bulk of a spacetime when gravity is dynamical, has been known at least since the time of Bryce DeWitt's papers in the 60's \cite{XX}. His observations were in the context of flat space quantum gravity where an S-matrix seems to be the only natural observable. This paucity of true bulk observables is now understood as automatic in AdS as well \cite{Witten}, where the degrees of freedom are naturally those of the boundary CFT. Given these observations, what is surprising is that experimentally we have discovered $local$ quantum field theory in a world that also contains gravity -- in a holographic Universe, it must surely be the case that the bulk locality we experience is only approximate. 

The trouble however, is that the approximations required to obtain bulk effective field theory from the ``microscopic" holographic degrees of freedom, are quite mysterious. Nonetheless, when the curvatures are small, the usual successful tests of the AdS/CFT correspondence in the supergravity limit may be viewed as evidence that bulk EFT-based calculations in a given background, do give reasonable results for many questions. Despite the mysterious origins of bulk locality, it is fair to say that the validity of bulk effective field theory around fixed classical backgrounds (eg., empty AdS) is believed to be non-controversial for large classes of questions.

But there do exist contexts where this expectation may be misleading. This is when we deal with very large timescales in the geometry, usually associated to the evaporation time (or the ``half-way" Page time) of a black hole. It is conceivable that that existence of large dimensionless parameters may qualitatively change our intuition, and perhaps it is not reasonable to view local bulk physics as sensible (even outside horizons!) in such situations. In this paper we will present a calculation that provides evidence that this worry is misplaced in the context of some of the recent questions regarding the information paradox. 

The backdrop of our work is the recent discussions of \cite{Penington, Almheiri, Mahajan1, Hartman1, Shenker} on the black hole information paradox \cite{Hawking, Page, Mathur, AMPS}. We consider a braneworld black hole coupled to a gravitating bath brane \cite{Karch2} in the ``wedge holography" \cite{Akal, Miao, Miao2} set up. This scenario (see Figure 1) is very similar to the one considered earlier in \cite{Mahajan1, Mahajan2, Geng}, but has the difference that the bath here has dynamical gravity. In the case of non-gravitating baths \cite{Mahajan1, Mahajan2}, computations of Ryu-Takayanagi \cite{RT} surfaces anchored to the bath showed that one can  formulate and resolve the information paradox unambiguously. This is the doubly holographic \cite{Mahajan1} version of  the island mechanism, for black holes on the brane. 

Our primary focus in this paper will be in the set up of \cite{Chamblin, Karch2}. The idea is to consider a black string in an AdS wedge bounded by two Karch-Randall branes \cite{KR}. This is a very nice set up because unlike some of the previously considered doubly holographic island systems \cite{Mahajan2, CK-critical}, it allows us to solve ODEs instead of PDEs when looking for islands, even in higher dimensions. The disadvantage is that the bath is $also$ a black hole, so it is quite different from the gravitating baths one expects in (say) the asymptotic regions of flat space black holes. The asymptotic region of the AdS wedge is at the defect which is in the ``middle" of the physical-brane/bath-brane system. So extracting general lessons from the system requires care. In \cite{Karch2}, the authors adopt the point of view that because of the dynamical nature of gravity on the bath brane, one should not anchor the RT surface anywhere specific on the bath\footnote{More concretely, they look for solutions that are Neumann on both ends.}. When you search for an RT surface on a compact geometry without anchoring it anywhere, it will straddle the bulk horizon\footnote{Recall that the RT surface associated to the entire boundary of a large global AdS black hole is its event horizon.}. This is the statement that the fully fine-grained Page curve of the combined system+bath is a constant, as opposed to the slide-shaped Page curve that emerges if one computes the entanglement entropy of just the system. The fine-grained result is a version of DeWitt's observation  that the exact observables of quantum gravity are holographic, and therefore have no support in the bulk \cite{Laddha}. In the second half of the paper, \cite{Karch2} considers RT surfaces anchored to the defect at the vertex of the wedge, where gravity is non-dynamical. When they do this, they find a Page curve which has an information paradox thanks to a Mathur-Hartman-Maldacena (MHM) surface \cite{Mathur, HM, Mathur2}, that gets resolved by an island. The authors argue that the emergence of the slide-shaped (unitarity-compatible) Page curve here is a consequence of the fact there is a tensor factorization in the dual defect CFT. The implied idea here seems to be that the Page curve one sees this way is not of particular interest for understanding unitarity of black hole evaporation --  degrees of freedom moving from one tensor factor to the other would of course $always$ lead to the unitarity-compatible Page curve\footnote{Our view is somewhat different. Both in the non-gravitational bath cases considered by \cite{Penington, Almheiri, Mahajan1, Mahajan2} as well as here, the $existence$ of a legitimate information paradox is a crucial point of interest. Adding the bath to a black hole exposes a paradox, but adding the bath to an ordinary quantum system does not. This is a strong suggestion that the bath and the associated tensor factorization are largely just a crutch here, the physics lies elsewhere. The new feature about \cite{Karch2} from this perspective is simply that the bath also has a holographic dual description. Note that whether there exists an approximate radial tensor factorization in (say) flat space quantum gravity, is a separate question. We claim that the answer to the latter question is also ``yes", but that is distinct from the point we are emphasizing in this footnote. }. 


In this paper, we will put the results of \cite{Karch2} in perspective by considering $Dirichlet$ boundary conditions on the  branes, when discussing potential RT surfaces. We have a few motivations for considering such boundary conditions. The first one is technical. As we will see, Dirichlet is a perfectly valid boundary condition allowed by the boundary terms arising in the variational problem. Therefore it is worthwhile investigating if there exists interesting solutions with these boundary conditions. Our second motivation is philosophic. Ruling out RT surfaces anchored to a gravitating bath, is unsatisfactory in our opinion, when done by fiat. It would be more instructive to allow Dirichlet as candidate boundary conditions $a$-$priori$, and explicitly check (a) whether they allow solutions, and (b) whether they have a natural limit where gravity decouples from the anchor. If such a limit exists and the physics is very different in that limit, $then$ it will be a strong argument that weakly gravitating anchors are fundamentally different from non-gravitating ones, as far as the information paradox is considered. In this paper, we will find that (a) Dirichlet solutions exist, and that (b) they do have a natural non-gravitating limit. Crucially, we will also find that the island physics is identical on a weakly gravitating anchor as it is on a non-gravitating one. In fact, the results in the second half of \cite{Karch2}  that we alluded to above, can be understood as the ``non-gravitating anchor limit" of our calculations. This gives a natural context to understand the slide-shaped Page curves there. Let us also emphasize that throughout our calculations, the graviton is massless, contrary to some suggestions in the literature that islands may not exist when the graviton is massless. 

Another way to state our philosophy, is as follows. In our view, not anchoring the surface anywhere is tantamount to treating the entire system+bath as one -- we are working with the holographic dual theory of the entire defect as one object. If one does not allow $any$ tensor factorization, then it is certainly true that there is no possibility of a conventional Page curve to begin with. But it is a dynamical question whether interesting (approximate) tensor factorizations exist. Our suggestion is that Dirichlet is a natural candidate for testing whether such a proposal can make  sense in gravity. We will find that it can. In contexts where bulk local physics is usually viewed as trustable (eg., outside the horizon), our calculation reveals that a conventional slide-shaped Page curve can be obtained from our Dirichlet calculation even when there is gravity on the bath. We can in fact go one step further, and show that the physics remains intact in the limit where gravity decouples from the anchor.

A final point we will make is that when defining entanglement entropy in a quantum system, one needs to make a $choice$ of the subsystem. The location of the Dirichlet condition we impose on one end of our RT surfaces can be viewed as the (approximate/semi-classical) gravitational analogue of such a choice. In particular, we are letting the other end of the RT surface land where it may via a more conventional Neumann boundary condition. This picks out the regions that are entangled with our chosen ``gravitational subsystem". 



 
We view our calculation as strengthening the case for a not-so-fine-grained, and therefore conventional (non-constant) Page curve, in settings where the sink gravitates. As emphasized in \cite{CK1}, a fully fine-grained approach to holographic information may not be the right path to a useful definition of the flat space Page curve. We will make more comments about approximate bulk locality, in the final section. In the rest of this section, we will discuss some previous works that considered closely related questions, and came to closely related conclusions. 

Information paradox and the Page curve for the flat space black hole were discussed in \cite{CK1}. An island mechanism was observed analogous to \cite{Penington}, upon introducing a Lorentzian cut-off/screen in the geometry. The significance of a Lorentzian screen for entanglement questions in holography (including a connection to quantum error correction) was suggested earlier in \cite{ACD}, see also \cite{Marolf}. The case for such a cut-off/screen was later emphasized in the ``central dogma" of \cite{review}, again for the flat space black hole -- the conclusion there was also that a tent-shaped Page curve is expected. In a 1+1 dimensional flat space context, where the graviton dynamics is trivial and therefore the distinction between a gravitating and a non-gravitating bath is less sharp, similar conclusions were also drawn earlier\footnote{Note that some of these 1+1 dimensional discussions are in the context of eternal black holes in flat space, see also the higher dimensional eternal BH discussion of \cite{I2}. To hold a black hole at fixed temperature in flat space, we need a radiation bath at infinity, which necessarily backreacts. One manifestation of this fact is that the Hartle-Hawking state of a flat space black hole is unstable \cite{Kay}. So these calculations are best viewed as suggestive, instead of being read literally. Adequately taking the backreaction into account leads to a black hole living in an FRW cosmology supported by radiation. This is precisely what one finds in the doubly holographic set up of \cite{CK-critical}, where the information paradox and its resolution are in terms of the $co$-$moving$ entanglement entropy.} in \cite{Thorlacius, Iizuka}.

The second case we will review deals with a black hole in a braneworld FRW cosmology, in equilibrium with radiation that is at the same temperature as the black hole. This in particular means that both temperatures are red-shifting at the same rate \cite{CK-critical}. A black hole  at the same temperature as the radiation supporting the cosmology may sound like a somewhat fabricated scenario. But note that the fine-tuning in temperature involved here is the same fine-tuning between bath and black hole engineered in any eternal black hole set up\footnote{This is in fact a natural set up for dealing with the eternal black hole when the bath is gravitating, see the previous footnote.}. The thing that makes this particular arrangement interesting is that when the radiation is viewed as comprised of deconfined large-$N$ matter, the system can be studied using a doubly holographic prescription where the brane lives in a black funnel geometry. The entanglement entropy can be studied via the RT surfaces in the underlying black funnel \cite{CK-critical}. This calculates a co-moving entanglement entropy on the braneworld. It was noted \cite{CK-critical} that it is precisely this object that keep tracks of black hole information (and the paradox) without being contaminated by the entanglement entropy of modes that become accessible due to cosmological red-shifting. Strong circumstantial evidence was presented in \cite{CK-critical} that the structure of RT surfaces in these funnel geometries is such that the island mechanism can resolve the information paradox in the co-moving entanglement entropy, despite the presence of massless gravitons. For the purposes of our discussions in this paper, the key observation regarding \cite{CK-critical} is that the boundary condition for RT surfaces on the brane there can be viewed as a co-moving Dirichlet boundary condition.  We will view that as the natural adaptation of the Dirichlet boundary conditions we adopt for the static geometries here, to a moving brane scenario. Note also the very recent work of \cite{Yasunori, Onkar} which also seem to be coming to similar conclusions as \cite{CK1, CK-critical} about the validity of gravitating baths.

In the next section, we will describe the wedge holography setting with a black string discussed in \cite{Karch2}. Our discussion is not self-contained; we will be fairly telegraphic except in emphasizing some of the points that will be important to us. In section 3, we will discuss the solution space of Dirichlet anchored extremal surfaces in this wedge geometry in detail, and observe that it naturally leads us to a picture where weak gravity at the anchor makes quite a bit of sense. This section contains the main technical details of the paper, as well as the features of the solution space. In the final section, we will present the case for a not-so-fine-grained Page curve when dealing with entanglement entropy in gravity and how this is the object that is significant for the information paradox in flat space black holes (as well as small black holes in AdS). Our discussion will focus on new arguments that have not been previously put forth in \cite{CK1}, and will also try to consolidate an overall picture.



\section{Black String Wedge Holography}

We consider a doubly holographic set up where two subcritical KR branes are embedded into an AdS black string wedge, see \cite{Karch2} for details. Our angle conventions follow their Figure 1, right panel. The left brane at $\theta_1$ is sometimes called the physical brane and the right brane at $\theta_2$ is the radiation/bath brane. We will often refer to them simply as the left or right brane, because what is viewed as system and what is bath is somewhat democratic in this system. We will see later that sometimes this is decided dynamically. The bulk metric in polar coordinate looks like \cite{Chamblin, Karch2},
\begin{equation}
ds^2=\frac{1}{u^2 \sin^2\mu}\left[-h(u) dt^2+\frac{du^2}{h(u)}+\vec{dx}^2+u^2d\mu^2\right],
\end{equation}
where $ h(u)=1-\frac{u^{d-1}}{u_h^{d-1}}$.

We can see that there is a planar $AdS_d$-Schwazschild black hole at each constant-$\mu$ slice with horizon at $u=u_h$. We set $u_h=1$ in the rest of the paper. Therefore both the branes have black holes and they are in equilibrium. We are interested in finding entangling surfaces that reside entirely outside the horizon. So we parametrize our RT surfaces as $u(\mu)$ and the corresponding action at $t=0$  becomes,
\begin{equation} \label{islandaction}
A=\int_{\theta_1}^{\pi-\theta_2} \frac{d\mu}{(u \sin \mu)^{d-1}}\sqrt{u^2+\frac{u'(\mu)^2}{h(u)}}.
\end{equation}
Varying this action we find 
\begin{equation}
\begin{split}
0=\delta A & =\int_{\theta_1}^{\pi-\theta_2}\frac{\delta A}{\delta u'}\delta u' d\mu +\int_{\theta_1}^{\pi-\theta_2}\frac{\delta A}{\delta u}\delta u d\mu\\
& =\left.\frac{\delta u}{(u\sin \mu)^{d-1}}\frac{u'(\mu)}{h(u)\sqrt{u^2+\frac{u'(\mu)^2}{h(u)}}}\right\vert^{\pi-\theta_2}_{\theta_1}-\int_{\theta_1}^{\pi-\theta_2} (\text{EOM}) \delta u d\mu
\end{split}
\end{equation}
The starting observation is that we do not necessarily have to use Neumann boundary conditions $u'(\theta)=0$ at both ends to satisfy the variational principle. Instead, we will impose Dirichlet boundary condition $u(\pi-\theta_2)=a$ at one end (``bath") and Neumann boundary condition $u'(\theta_1)=0$ at the other (``physical brane"). This is a perfectly acceptable class of solutions of the variational problem such that the boundary variation of $A$ vanishes.

Though the discussion can be generalised to any $d$, we focus on $d=4$ for illustrating our goal. Note that the dimensionality is high enough to incorporate non-trivial metric dynamics. It is also worth emphasizing that the spectrum of excitations of this system contains a massless graviton. Imposing the boundary conditions above we solve the following equation of motion,
\begin{equation}
\begin{split}
& u''(\mu )-\frac{(d-1) \cot (\mu ) u'(\mu )^3}{u(\mu )^2 h(u(\mu ))}+d u(\mu ) h(u(\mu ))+\frac{(d-3) u'(\mu )^2}{u(\mu )}-(d-1) \cot (\mu ) u'(\mu )\\
&-\frac{u'(\mu )^2 h'(u(\mu ))}{2 h(u(\mu ))} -2 u(\mu ) h(u(\mu ))=0
\end{split}
\end{equation}

We will also be interested in finding Mathur-Hartman-Maldacena (MHM) surfaces. We parametrize these surface as $\mu(u)$ and impose Neumann boundary conditions at the horizon. The corresponding action takes the following form at $t=0$,
\begin{equation} \label{HMaction}
A=\int_{u_0}^{u_h} \frac{du }{(u \sin \mu)^{d-1}} \sqrt{\frac{1}{h(u)}+u^2 \mu'(u)^2}.
\end{equation}
To find the equation of motion we first vary the above action and keep track of boundary terms,
\begin{equation}
\begin{split}
0=\delta A & =\int_{u_0}^{u_h}\frac{\delta A}{\delta \mu'}\delta \mu' du +\int_{u_0}^{u_h}\frac{\delta A}{\delta \mu}\delta \mu\, d u\\
& =\left. \frac{\delta \mu}{(u\sin \mu)^{d-1}} \frac{\mu'(u) u^2}{\sqrt{\frac{1}{h(u)}+u^2 \mu'(u)^2}}\right\vert_{u_0}^{u_h}+\int_{u_0}^{u_h}(\text{EOM})  \delta \mu \,du
\end{split}
\end{equation}
 The corresponding equation of motion is 
 \begin{equation}
 \mu''(u)+\frac{3 \cot \mu(u)}{u^2-u^5}+\frac{(2+u^3)\mu'(u)}{2u(u^3-1)}+3\cot \mu(u)\mu'(u)^2+2u(u^3-1)\mu'(u)^3=0.
 \end{equation}
\begin{figure} [h]
 \includegraphics[width=\linewidth]{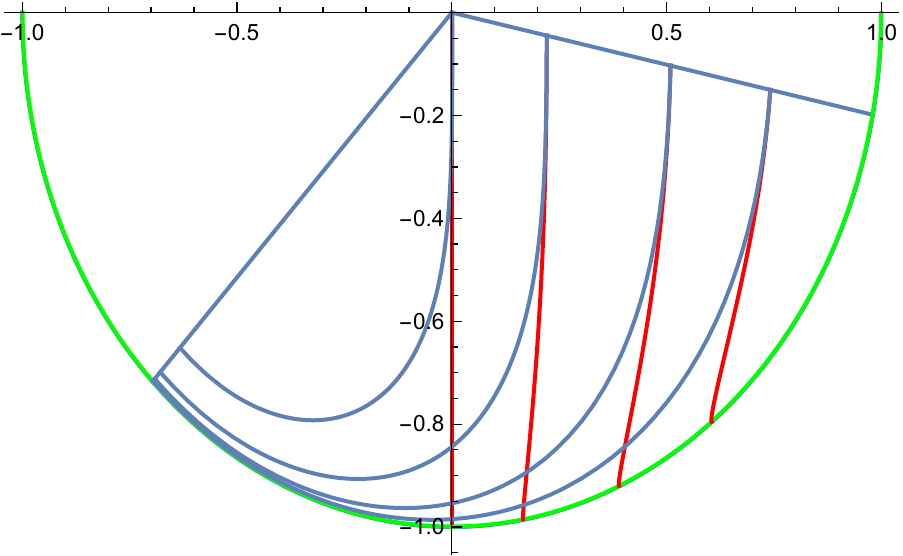}
    \caption{A sample of Dirichlet anchors from the bath brane for $\theta_1=0.8, \theta_2=0.2$. MHM surfaces are red, island surfaces are blue.}
    \label{simple}
\end{figure} 
Notice that $\mu=\frac{\pi}{2}$ is the simplest solution of the above equation and that is the MHM surface that is anchored to the defect. This was the object of interest in \cite{Karch2}. 

We will be interested in Dirichlet anchoring the RT surfaces at arbitrary locations on the branes. This means that we will have to solve the above equations of motion numerically. We will study the detailed phenomenology of these curves\footnote{We will often use the words ``curve" and ``surface" interchangeably.} in the next section, but let us make one of our key observations with an illustrative example, shown in Figure \ref{simple}. Two classes of curves that emerge from the Dirichlet anchors on the right (weaker gravity) brane are shown. The red curves that fall into the horizon are the MHM curves. The curves that land on the physical brane are the island curves.

After finding the numerical solutions of the various curves, we compute the corresponding areas of the island surfaces and the MHM surfaces using  \eqref{islandaction} and \eqref{HMaction}. Note that except when the anchor is at the defect, all our results are manifestly finite.  When the brane angles are below the so-called Page angle \cite{Karch2}, MHM are always the dominant (ie., minimal) curves at $t=0$. But the blue (island) curves that land on the physical brane with a Neumann boundary condition are finite, and resolve the information paradox at late times, with a non-trivial Page curve. 

For this class of curves, the structure of island physics near the defect is isomorphic to that at the defect found in \cite{Karch2}. In fact, this remains true for all the various cases we will discuss in the next section -- this is what we mean by the statement that the limit to the non-gravitating anchor is continuous. In the next section, we will see a whole bestiary of curves, and discuss some of their implications. 
   \begin{figure}[hbt!]
  \begin{subfigure}[b]{0.48\linewidth}
    \includegraphics[width=\linewidth]{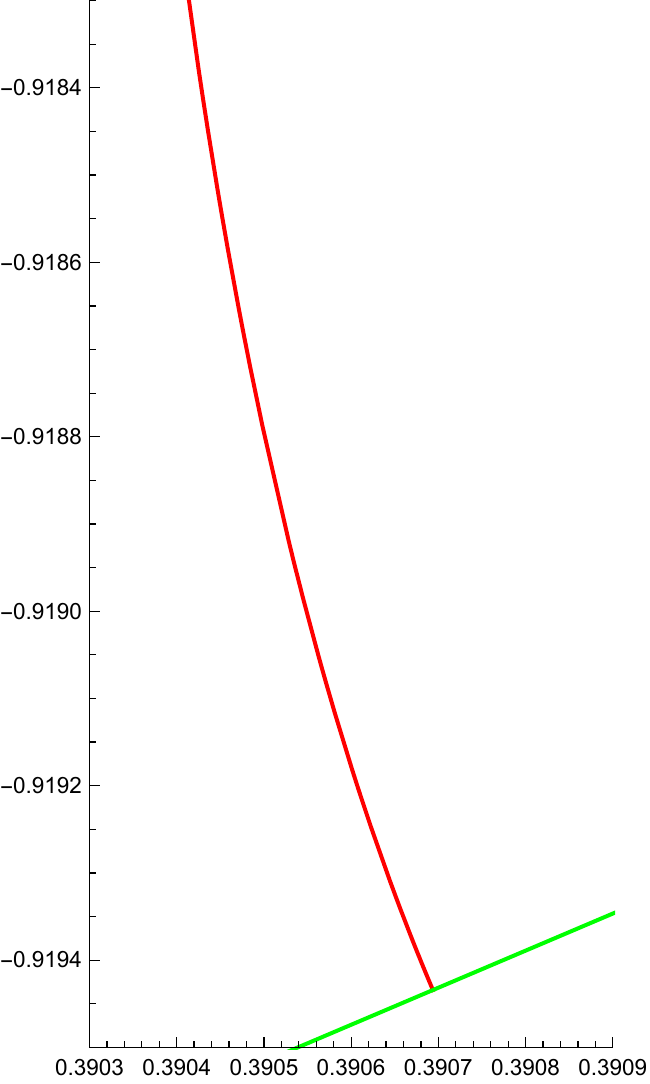}
    \caption{$\pi/2+0.4$}
  \end{subfigure}
  \begin{subfigure}[b]{0.48\linewidth}
     \includegraphics[width=\linewidth]{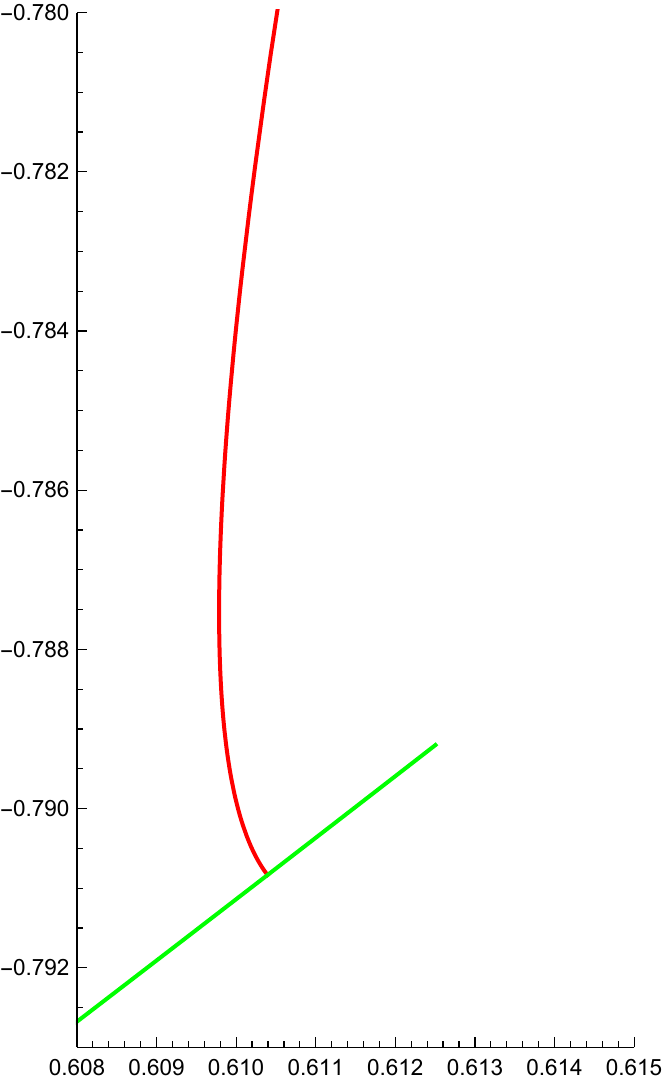}
    \caption{$\pi/2+0.65$}
   \end{subfigure}
   \caption{The zoomed in plot near the horizon for MHM surfaces when $\theta_2=0.2$. The captions denote the angle at which the surfaces reach the horizon (normally).}
   \label{im:$1d$_2dising}
\end{figure} 

The resolution of the MHM curves near the horizon in Figure \ref{simple} is not sufficient to make it obvious that those curves are hitting the horizon normally. So we show the near-horizon region of these curves in a zoomed-in Figure \ref{im:$1d$_2dising} to emphasize that indeed everything is as it should be, and the boundary condition at the horizon is Neumann. 

\section{Ryu-Takayanagi Phenomenology}

In this section, we outline the detailed phenomenology of the extremal/RT surfaces in the wedge geometry with the two branes. They result in information paradox in many of the configurations, but islands always emerge to salvage the day. 

Specifically, we anchor the extremal surface 
to one of the two branes via a Dirichlet condition, and we vary (a) the location of the anchor, and (b) the angles of either brane. We will also allow the possibility that the anchor can be on either brane. This scans through the entire spectrum of allowed curves in this system with one Dirichlet end and one Neumann end. Our philosophy is that in a dynamical set up involving gravity, it is more meaningful to let the dynamics decide what is the natural choice of bath/radiation once we pick a Dirichlet location, rather than pick the bath brane beforehand. This leads to some interesting physics as gravity gets stronger on the branes and as we move deeper into the bulk.

\subsection{Curve Phenomenology on a Single Brane}

\begin{figure} [h]
 \includegraphics[width=\linewidth]{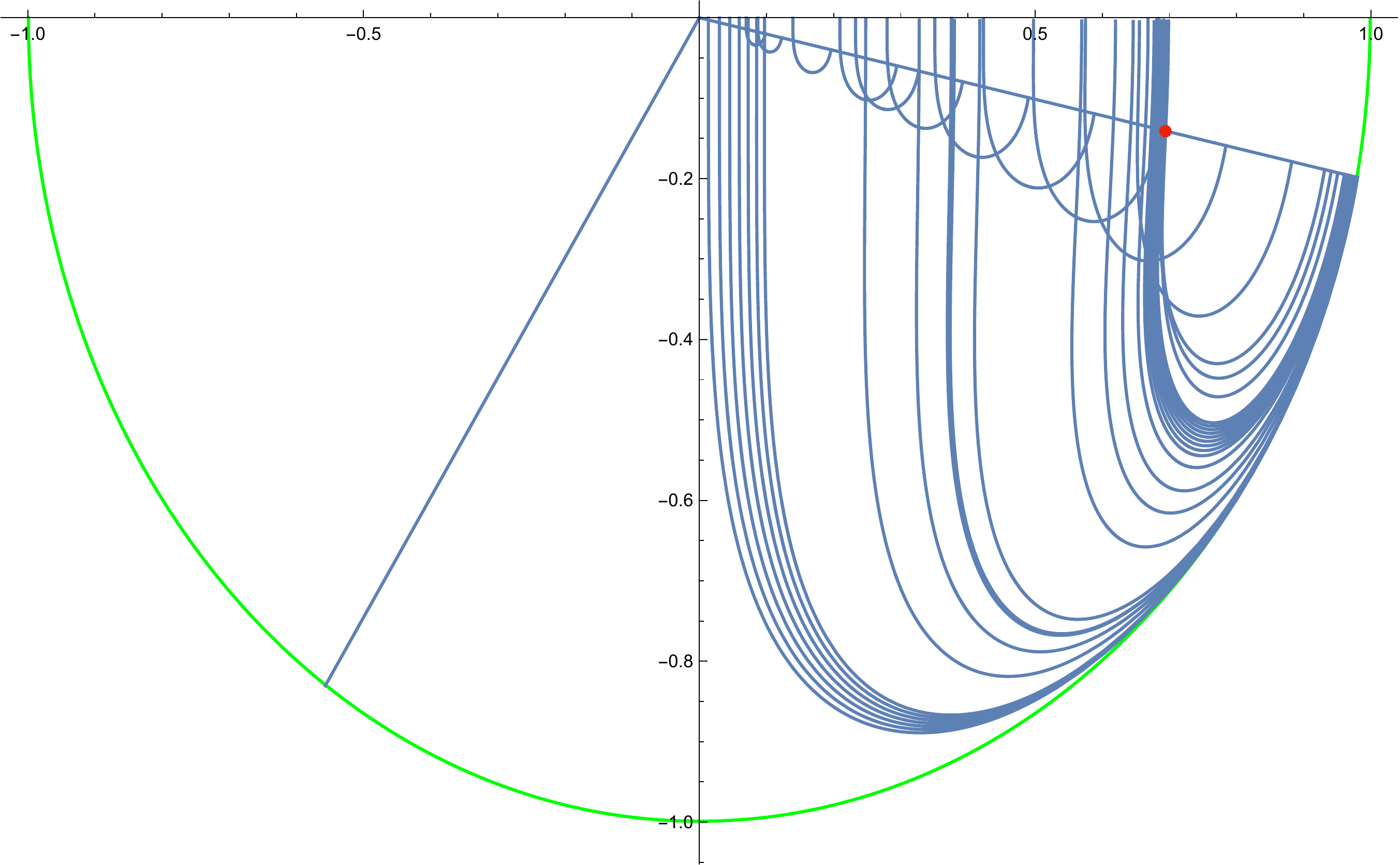}
    \caption{A collection of curves that start and end on a single brane. The Neumann ends are shown as ending on the brane, the Dirichlet ends are extended to the boundary of the AdS geometry. The red dot is the Dirichlet wall.}
    \label{single-brane}
\end{figure}

Our discussion splits up into various sub-cases. Before we discuss these cases, it is useful to discuss the case with a single brane, where we look for curves that start on the brane with a Dirichlet anchor. There is a rich structure to these curves, and they will be crucial for us in understanding the more complicated situation involving two branes. The first observation is that as long as the angle of the brane is less than the critical angle $\theta_c$, and as long as the Dirichlet anchor is close enough to the defect, the curves that start Dirichlet can always end Neumann on the same brane. In fact, there exist $two$ curves that end with Neumann boundary condition, that start on the brane with Dirichlet. A typical structure of these curves is shown in Figure \ref{single-brane}. Of the two curves, the dominant one (that is, the one with the minimal area) is the only one that will affect our discussions\footnote{MHM surfaces are the truly dominant surfaces at $t=0$ for brane angles that are below $\theta_p$, but in our discussions here, we are emphasizing the relative dominance between surfaces that start and end on branes.}. When we talk about a curve starting and ending on the same brane without further qualifications, this dominant curve among the two will be the one we will typically have in mind. 

A crucial observation about the Dirichlet anchored curves that end Neumann on the same brane is that {\em they do not exist}, once the Dirichlet anchor is more than a certain distance (that depends on the angle of the brane) from the defect. This location we call the ``Dirichlet wall" (or D-wall for short) and is marked by a dot on the branes in our figures. Beyond the D-wall, curves can still be Dirichlet anchored to the brane, but they can only end on another brane (assuming there is a second brane present) if we demand Neumann boundary conditions on the other end. We will view the D-wall as the transition from weak gravity to moderately\footnote{``Moderately", because we will see that it still makes sense to anchor curves there. Islands and Page curves still work. But the $structure$ of the curves is qualitatively different from near the defect.} strong gravity on the brane as far as the anchor is considered. 

The D-wall is always inside\footnote{ie., closer to the defect than the critical anchor.} the {\em critical anchor}. The latter is the point on the brane from which a curve with Neumann boundary conditions, reaches the defect \cite{Karch2}. This means that once the brane angle crosses $\theta_c$ so that the critical anchor has gone into the defect, the D-wall is also at the defect. So none of the curves that start Dirichlet on the brane can end Neumann on the same brane, if the brane angle is greater than $\theta_c$. They all have to go to another brane (if there is one) so that they can land on it with a Neumann boundary condition.  

Note that from any Dirichlet anchored point on one of the branes, there is $always$ an extremal curve that lands with Neumann boundary conditions on the horizon. This is the Mathur-Hartman-Maldacena (MHM) surface, and the fact that it cuts the horizon means that its area increases unboundedly with time. This is a version of the information paradox \cite{Mathur, HM} and if this were the only extremal surface starting from the Dirichlet anchor, we would have a problem. But we will see that there are $always$ other extremal surfaces that start on the same Dirichlet anchor and end either on the same brane or on the other brane with a Neumann boundary condition. Since these surfaces are always of finite area\footnote{There is a slight subtlety for the special case when the Dirichlet end of the curve starts at the defect -- the IR divergence of the AdS boundary makes the areas infinite, but the divergences in the MHM curves and the island curves are identical, so they cancel in the differences. So it is meaningful to compare them.}, these island surfaces will be able to resolve the information paradox. One of the points we will clarify in the ensuing discussions is the question of which among the multiple possible island surfaces (when more than one exists) is the minimal RT choice. 

As long as the angles involved are less than $\theta_p$, the MHM surfaces are the dominant surfaces emanating from a Dirichlet anchor at $t=0$ (this is true, even if there are two branes). This means that in all such cases we will have a genuine information paradox, and the island is what will resolve it. Once the left brane angle goes above $\theta_p$, the MHM surface becomes subdominant for Dirichlet anchors that are close enough to the defect on the right brane\footnote{There is a bit more structure to this story, which we will describe when we discuss the cases where the brane angle is bigger than the Page angle.}. This means that the dominant surface is the island surface already from $t=0$ and therefore there is no information paradox, and the Page curve remains flat throughout. 

With these preliminary comments, we are ready to discuss the detailed structure of the RT surfaces and the island mechanism, with both branes in the wedge. 

\subsection{Curve Phenomenology with the D-Bath}

\subsubsection{$\theta_p > \theta_1 > \theta_2$}

\begin{figure} [h]
 \includegraphics[width=\linewidth]{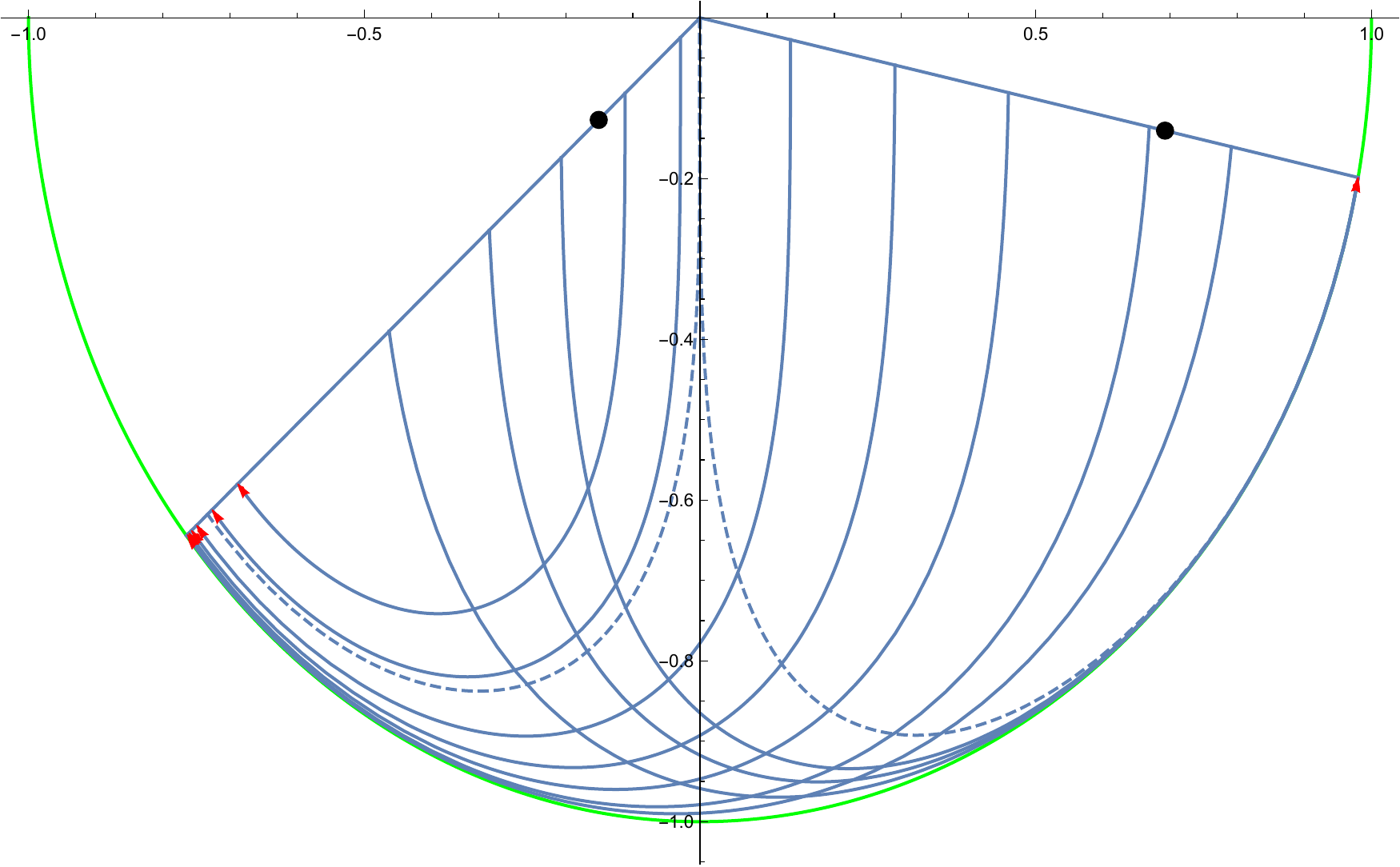}
    \caption{The space of curves with $\theta_1=0.7, \theta_2=0.2$. Neumann ends are denoted by red arrows, Dirichlet ends are unmarked. The black dots denote the D-walls, the dashed lines are the curves that connect the critical anchor to defect.}
    \label{Case1}
\end{figure}

We will first discuss the case where the physical brane is at a large angle\footnote{We will take the angle to be smaller than the Page angle, $\theta_p$, here. If it is above the Page angle, the MHM surface in a neighborhood of the defect on the smaller angle brane are sub-dominant to the island surfaces and there is no information paradox for those anchors. This does not substantively change the discussion, but we will deal with it as a special case later.} while the bath brane is at a small angle (eg., $\theta_1 = 0.7$ and $\theta_2 = 0.2$). In this case, the structure of the plots is presented in Figure \ref{Case1}. A key observation (and this is universally true, irrespective of the angles of the branes) is that as long as the Dirichlet anchor is close enough to the defect, the nature of the island physics is identical to that when the anchor is $at$ the defect. This is true regardless of whether the anchoring is on the left brane or the right brane\footnote{Without loss of generality, in our discussions we will set the bigger angle brane to be the left brane.}. This is a key indication that there is continuity between the weakly gravitating situation we consider, and the non-gravitating anchor considered in the second half of \cite{Karch2}. Weak gravity on the anchor, does {\em not} fundamentally change the island physics, as long as the island is close enough to the defect.  

When the angle difference between $\theta_1$ and $\theta_2$ is sufficiently large, we find that for $all$ Dirichlet anchors on the right brane, the dominant curve is the one that ends on the left brane\footnote{By dominant here, we mean among the curves that end on the branes. As mentioned previously, at early times the MHM surfaces is the truly dominant (ie., minimal) surface for all the Dirichlet anchors as long as the angles are below $\theta_p$.}. This is true even if the Dirichlet anchor is arbitrarily far from the defect. This means that on the right brane, the physics associated to the D-wall does not play a substantive role (even though we have marked it on the figure). This is consistent with the fact that a highly gravitating left brane makes the right brane effectively non-gravitating by comparison. 

This is not the case for the D-wall on the left brane. As long as the Dirichlet anchor is on the left brane and is closer to the defect than the D-wall, we find that the dominant island curves end on the $left$ brane itself\footnote{Note that this is necessary for the existence of a smooth limiting case for the Dirichlet anchor, as the anchor moves to the defect.}. This is natural, because as we will see throughout our discussion here, the curves are attracted to regions of higher gravity. The trade-off between the attraction due to the two branes depending on their angles is what largely sets the phenomenology of the curves.    There are two factors that contribute to how ``strong" or ``dynamical" gravity is. The first factor is the angle of the brane -- the steeper the angle of the brane the larger the $G_N$ associated to it. This is why it is somewhat reasonable to view the larger angle brane as the physical brane and the other one as the bath brane. But there is a subtlety to it, because the physics is also affected by a second factor - how deep into the bulk is the location on the brane where we are placing the Dirichlet anchor? The deeper into the bulk we go, the more ``dynamical" we can expect gravity to be, and corresponding to that we find that (if we only had one brane) we cannot place a Dirichlet anchor and expect the brane to land on itself with a Neumann boundary condition. This is the origin of the D-wall, that we discussed in the previous sub-section. 

When there are two branes and both are gravitating, this becomes more interesting. We find that we can in fact place Dirichlet boundary conditions beyond the D-wall, as long as we allow the Neumann boundary to land on the $other$ brane. A very heuristic interpretation of this is that since both branes are gravitating and holographic, both are trying to ``repel" their degrees of freedom into each other through the boundary defect -- and the place where they find balance is where we find an acceptable Dirichlet anchor. Note that when the Dirichlet anchor was before the D-wall of the left brane, the bath region is the region to the right of the anchor (including the right brane), and the island was the region to the left of the Neumann end of the curve. But once the anchor is farther from the defect than the D-wall of the left brane, since the Neumann end of the curve is now on the right brane, the bath is on the left brane and the island is on the right brane. In other words, what is the bath is decided dynamically, once we pick a Dirichlet location -- we find this to be extremely natural, since both branes contain gravity (and indeed, horizons!).

One curiosity we observe is that for the values of the angles that we have checked, the island region is smaller in size\footnote{This is a statement about coordinate ranges, so should not be taken too literally. But since it does have the smell of the correct physics, we will simply note it here.} than the bath region. This is what one would expect if the system was a gravitating brane coupled to a non-gravitating bath. It will be interesting to understand if this is universally true and to study the contexts where it fails, if at all it is not universal. This will take a more refined scan of the parameter space than what we have done in this paper. 

The present case, is in many ways the closest analogue in this system, to the situation where a gravitating brane was coupled to a non-gravitating bath \cite{Mahajan2}. The picture we find is structurally quite parallel, except for the fact that bath-vs-island choice is made $dynamically$. A natural philosophy regarding this is that the relative locations of the Dirichlet and Neumann ends of the brane keep track of which regions of the gravitating system contain degrees of freedom that are entangled with each other. We will make some speculative comments about this in the concluding section.

\subsubsection{$\theta_p > \theta_1 \sim \theta_2$}

As the angle difference between the left and right branes decreases, we find that there is some interesting structures that emerge. These structures do not substantively affect our main takeaway message regarding the parallels between gravitating and non-gravitating baths, but we will present them here for completeness. 

\begin{figure} [h]
 \includegraphics[width=\linewidth]{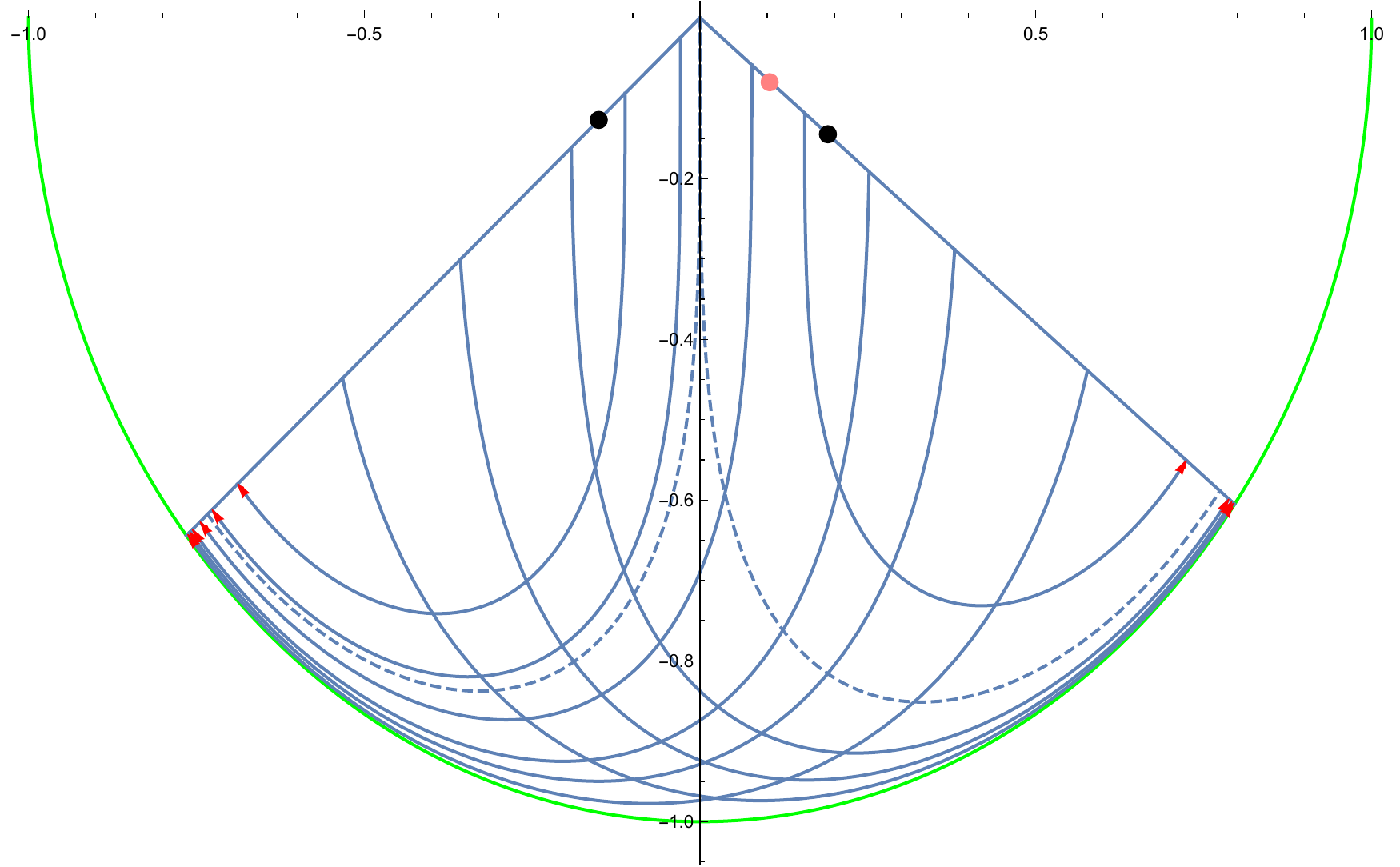}
    \caption{The space of curves with $\theta_1=0.7, \theta_2=0.65$. Neumann ends are denoted by red arrows, Dirichlet ends are unmarked. The black dots denote the D-walls, the dashed lines are the curves that connect the critical anchor to the defect, the pink dot is the Dirichlet critical point.}
    \label{Case2}
\end{figure}

The first observation is that as the angle of the right brane gets closer to that of the left brane (eg., $\theta_1=0.7$ and $\theta_2=0.65$), the right brane itself starts attracting the Dirichlet anchored curves that start from it. See Figure \ref{Case2}. So some of the curves that start from the right brane (obviously, closer to the defect than its D-wall) end up landing on the right brane, and these become the dominant curves\footnote{Again, we are not discussing the relative dominance of the MHM curves here, which are the truly dominant curves at $t=0$.}. The right brane curves that are dominant are the ones close to the right D-wall. As the anchoring point moves closer to the defect from the D-wall there is a phase transition at a location on the right brane which we will call the {\em Dirichlet critical point}. For Dirichlet anchors that are closer to the defect than the Dirichlet critical point, the dominant curves from the right brane are the ones that end on the left brane. This then continues to be the case as we keep moving our anchor all the way to the defect and then up through the left brane to the left D-wall, as is familiar from the previous case we discussed. The physics again proceeds in parallel to the previous case once we cross the left D-wall, and we find that all the dominant curves from the left brane now end on the right brane. 

A degenerate case worth a brief mention is when $\theta_2$ becomes equal to $\theta_1$. When this happens, the Dirichlet critical point coincides with the defect, and therefore $all$ the Dirichlet anchored curves\footnote{Let us again emphasize that we are excluding the MHM surfaces.} that start from the left brane between the D-wall and the defect, end on the left brane itself. Since $\theta_1$ is equal to $\theta_2$, a symmetric story appears on the right brane as well. This is a fine-tuned case which may be interesting for some purposes, but not for us here. As $\theta_2$ continues to increase, the Dirichlet critical point moves to the left brane, and the story we narrated in the previous paragraph continues with the left and right branes interchanged. As $\theta_2$ increases substantially above $\theta_1$, the Dirichlet critical point moves into the left D-wall, and the story reduces to the previous case we discussed, but again with right and left branes interchanged.

\subsubsection{$\theta_1 > \theta_p$ } 

New cases of some interest occur when the brane angles start going above the Page angle. Here again there are a few cases which can be dealt with separately. Let us first consider the case where $\theta_1$ is bigger that $\theta_p$. Note that $\theta_p$ is the angle at which the MHM surface from the defect becomes subdominant to the island surface connecting the critical anchor and the defect, already at $t=0$. This means that the Page curve for brane angles above $\theta_p$ are flat for the defect anchor; the island surface is the one that computes the entropy and its area stays constant. 

For $\theta_1 > \theta_p > \theta_2$, we find that it is not just the defect anchor that has a flat Page curve. For a finite range of Dirichlet anchors on the right brane near the defect, we have flat Page curves. The boundary of the region on the right brane where the Page curves start becoming non-trivial, we call the {\em boundary of triviality}. The boundary of triviality expands out to bigger and bigger regions, as $either$ of the angles $\theta_1$ or $\theta_2$ is further increased. Nothing qualitatively changes for the right brane discussion, even after $\theta_2$ becomes greater than $\theta_p$. Of course, when this happens, the left brane starts having a region of trivial Page curves for curves that are anchored close enough to the defect. Until then, the left brane anchors all had non-trivial Page curves. The left and right brane stories are completely symmetric. 

In principle we could also separately consider the case where the brane angles are bigger than the critical angel $\theta_c$ which is somewhat bigger than $\theta_p$. But note that from our point of view this is simply the situation where the critical anchor has already gone into the defect and therefore simply a particular case of our discussion above. The only notable difference is that the surfaces that start at a Dirichlet anchor close to the defect on the right brane and end on the left brane (or vice versa) will end at a vanishing location (ie., on the defect itself) on the left brane, if the latter brane is at an angle bigger than $\theta_c$. This is because above the critical angle, the critical anchor has gone into the defect. These curves were called ``tiny" island surfaces in \cite{Karch2}, our Dirichlet approach therefore provides a simple understanding for their origin. But as far as the qualitative physics is considered, they are simply a continuation of the discussion we had in the above paragraphs as the angle(s) steadily increase(s) beyond $\theta_p$. It may be of some interest to compute entropies explicitly for a refined scan, and make a quantitative statement about its continuity properties. But we will not undertake it here. We have checked that the above discussion applies even as the angles get quite close to $\pi/2$. 



A key feature of the discussion above is that the space of curves we have noted has a very nice continuity structure as we tune the Dirichlet anchor location, or either of the brane angles. The emergence of new phases and changes of dominance of the phases have a very clear and simple understanding. This provides us considerable confidence that the overall picture of the space of curves above is correct, even though the construction is numerical and finer scans of the parameter space may be of some interest. 


\section{The Not-so-Fine-Grained Page Curve}

Despite the lack of exact bulk locality in the presence of gravity \cite{XX}, there has been a long history of calculations, where introducing a cut-off in the bulk has lead to reasonable results, especially for computing gravitational  free energies and (entanglement) entropies. Perhaps the original example of this is the famous work of Gibbons and Hawking \cite{GH}, where a Euclidean cut-off was introduced. One of the implications of \cite{ACD, CK1, review} would be that a cut-off in Lorentzian signature is also of utility when discussing entanglement questions in flat space holography\footnote{Euclidean calculations involving replica wormholes have recently been argued \cite{Hartman1, Shenker} to shed light on Lorentzian questions about the information paradox.}. Since the radial direction is expected to be related to energy scale in the holographic description, it seems plausible that the relevant (approximate) tensor factorization has something to do with energy scale. Understanding this precisely would of course be a major step forward in our understanding of bulk locality in holography. In what follows, we will make some qualitative comments regarding the possibility of such a ``not-so-fine-grained" Page curve \cite{CK1, Vyshnav}\footnote{The implication being that in a fully fine-grained description, the exact holographic degrees of freedom of gravity would be living at infinity, and so we can only expect a constant Page curve \cite{Karch2, Laddha}.}. A handful of observations which support such a point of view were already presented in \cite{CK1}, here we will try to make some new comments.

\begin{itemize}
\item Our Dirichlet prescription here can be viewed as the doubly holographic analogue to the Gibbons-Hawking idea of fixing field values at the cut-off. It would be interesting to see if a more natural formulation would be to work with sources at the boundary instead of boundary values, as suggested in \cite{Buddha}.
\item One way to motivate that a non-trivial Page curve can exist in flat space is to start with the usual near-horizon limit argument for the AdS/CFT correspondence. The near-horizon limit is a low energy limit, and by taking it, we are decoupling the brane (or black hole) degrees of freedom from those in the asymptotic region. But it is believed that the system admits holographic descriptions, both ``before" and ``after" the limit\footnote{``Before", because we believe that flat space quantum gravity is holographic. ``After", is just the usual AdS/CFT correspondence.}. The fact that there exists a holographic description even after the near-horizon limit, is suggestive. This enables us to view stepping back from the near horizon limit as analogous to coupling the ``after" holographic system to a sink, like in \cite{Penington, Almheiri}. Of course, these arguments are not watertight because only extremal/charged/supersymmetric black holes admit simple near-horizon limits, but the fact that black holes have area-scaling of entropy we believe is a strong suggestion that this picture is approximately correct. After all, this was one of the original motivations for speculating that gravity is holographic. The island prescription invokes the non-locality associated to holography in the ``after" description, coupled to a bath -- while the fully fine-grained constant Page curve of \cite{Laddha} is in the ``before" description. Note that these two separate notions of non-locality/holography are necessary for the island picture to make sense in flat space. 
\item Continuing the above discussion, let us also note that the usual ``derivation" of the AdS/CFT correspondence  explicitly identifies a tensor factorization in the {\em bulk} degrees of freedom. The near horizon limit leads to one tensor factor that captures horizon physics (which can be thought of as AdS$_5$ string theory or ${\cal N}=4$ SYM in the case of D3-branes in flat space) and a decoupled sector of free gravitons in the asymptotic region. Despite the bulk-localized nature of the near-horizon region, the result is beleived to capture gauge invariant information. A natural explanation for this exists -- the near-horizon limit is a low energy limit due to the redshift  at the horizon and the asymptotic Hamiltonian constraint is not resolved enough to detect fluctuations there. It is natural to suspect that a finite temperature version of this where the horizon region and the asymptotic region are coupled, is the context of the flat space information paradox.  
\item The cut-off of \cite{CK1, ACD} and the central dogma of \cite{review} treat black holes as approximately localized systems in the bulk. Note that Hawking's calculation of Hawking radiation is also a bulk-local calculation. To argue that one can extract information from the black hole to get a conventional Page curve \cite{Page}, perhaps one needs to find\footnote{We thank S. Raju for this comment.} a limit where this information is localized in the black hole tensor factor in same sense that it is, in a local theory. The trouble is of course that in the ``local" limit where gravity turns off ($G_N \rightarrow 0$) the black hole information/entropy diverges. A complete resolution to this problem is not known to us, but one observation that was made in \cite{CK1} was that unlike in say ordinary non-gravitational thermal systems, a small black hole in AdS space may come with a small parameter, which can substantively change this naive reasoning. This is because there is some evidence that a small black hole \cite{Berenstein, Hanada} should be viewed as a deconfined configuration of $M \times M$ sub-matrices in a holographic theory of $N\times N$ matrix fields. This brings up a dimensionless small parameter $M/N$ into the problem.    
\item Let us also address one criticism that has been raised in the refereeing process of this paper -- that all we have done in this paper is to simply identify some ``minimal surfaces". It is certainly true that if one strips our calculations off their context, all we have done is demonstrate the existence of certain RT surfaces. But the point is that these island carrying RT surfaces need not have existed, and without them there would have been an information paradox. But these RT surfaces do exist, whenever there is a paradox-making MHM surface in the bulk. The key point is that this is {\em irrespective} of whether there is a black hole on the boundary/brane like in the recent works, or if the boundary only contains an ordinary thermal system like in the original work of \cite{HM}. The fact that the bulk-computed Page curve remains the {\em same}   
irrespective of whether the boundary system is an ordinary (localized) thermal system or a black hole coupled to a gravitating/non-gravitating bath, is a strong suggestion in our mind that the conventional (non-constant) Page curve is indeed of significance in understanding the information paradox. Instead of denying the naturalness of these results, in our mind a more fruitful question is -- {\em what} is a suitable way to define bulk observables and observations? How can one reconcile them with the fact that the bulk is in some ways a gauge artefact?
\item Finally, let us point out another circumstantial piece of evidence for believeing that Dirichlet boundary conditions may be reasonable boundary conditions on gravitating braneworlds\footnote{We thank Dominik Neuenfeld for this comment.}. It was shown in \cite{Dominik} that extremizing the bulk RT surface with Neumann boundary conditions is equivalent to taking any brane entangling surface and extremizing the island functional. The latter of course only makes sense if we are free to choose an entangling surface on the brane, which directly points to Dirichlet conditions.
\end{itemize}

A key question we have $not$ addressed in this paper is the following -- what is the (approximate) tensor factorization associated to the entanglement entropies we are apparently computing? In what sense is this tensor factorization approximate?  These are closely tied to the question of how the degrees of freedom are organized in gravity in a manner that illuminates the origin of the holographic direction. Why does the holographic direction ``feel" at least for some questions, like just another spacetime coordinate? Answering this question can lead us to a suitable $bulk$ definition of the entanglement entropy, which presumably must take into account the organization of degrees of freedom according to energy scale. It also seems plausible that bulk gauge invariance (which is entirely invisible from the boundary) may play an important role. These considerations could provide a natural notion of coarse-graining associated to holography. It would be interesting if the rich phenomenology we have uncovered for the Dirichlet anchored surfaces here is useful in making progress on this question. 

Before we conclude, let us make some speculations about the nature of the coarse-graining/approximations involved in defining a suitable gravitational entanglement entropy. It seems plausible to us that the screen/cut-off/Dirichlet anchor is capturing the scale of the coarse graining \cite{CK1}. Note here that the conjectured holographic correspondence in flat space is a UV/UV correspondence \cite{Sam} as emphasized in \cite{CK1}. A direct definition of a bulk entanglement entropy of this type is likely a hopeless task at this stage in our understanding of string theory. But it has already been noted \cite{ACD, CK1} that  extremal surfaces anchored to such cut-off surfaces in Minkowski space satisfy features like strong-subadditivity\footnote{In work that is unpublished \cite{Vyshnav2}, we have demonstrated that other (holographic) entanglement inequalities also hold.}. In fact, it was noted in \cite{ACD, CK1} that more is true -- the structure that emerges has precisely the union/intersection structure that motivated the introduction of quantum error correction in AdS space \cite{ADH}. These works deal with genuinely Lorentzian settings. Let us also point out that the nature of the arguments in \cite{Headrick} which demonstrated subadditivity in a (Euclidean) holographic setting, were based on general features of the screen and the bulk, and does not rely on the details of AdS/CFT. 

We conclude by noting some of the latest works on the information paradox, as a point of entry into the recent developments \cite{Kundu, Malvimat, Choudhury, Wang, Ka, Karan, Arpan, Deng}.

\section{Acknowledgments}

We thank Andreas Karch, Vyshnav Mohan, Dominik Neuenfeld, Jude Pereira, Suvrat Raju and Sandip Trivedi for discussions.

\appendix

\end{document}